\documentstyle{article}
\begin{document}
\def\spose#1{\hbox to 0pt{#1\hss}}
\def\simlt{\mathrel{\spose{\lower 3pt\hbox{$\mathchar"218$}}
     \raise 2.0pt\hbox{$\mathchar"13C$}}}
\def\simgt{\mathrel{\spose{\lower 3pt\hbox{$\mathchar"218$}}
     \raise 2.0pt\hbox{$\mathchar"13E$}}}
\def\etal{{\it et al.\ }}
\def\s{{\rm\thinspace s}}
\def\erg{{\rm\thinspace erg}}
\def\ergps{\hbox{$\erg\s^{-1}\,$}}
\def\km{{\rm\thinspace km}}
\def\kmps {\hbox{$\km\ \s^{-1}\,$}}
\def\ref{\par \noindent \hangindent=3pc \hangafter=1}

\title{Astrophysical Evidence for Black Holes\thanks{To appear in:
`Black Holes and Relativity', ed. R. Wald. Chandrasekhar Memorial Conference, Dec. 1996.}}
\author{Martin J. Rees\\
Institute of Astronomy,
Madingley Road,
Cambridge, CB3 OHA}

\maketitle

\begin{abstract}   
The case for collapsed objects in some X-ray binary systems continues
to strengthen. But there is now even firmer evidence for supermassive
black holes in galactic centres.  Gravitational collapse seems to have
occurred in the centres of most newly-forming galaxies, manifesting
itself in a phase of quasar-like activity (which may be reactivated
later).  These phenomena (especially the gas-dynamical aspects) are
still a daunting challenge to theorists, but there is `cleaner'
evidence, based on stellar dynamics, for collapsed objects in the
centres of most nearby galaxies.  The current evidence does not tell
us the spin of the collapsed objects -- nor, indeed, whether they are
described by Kerr geometry, as general relativity theory
predicts. There are now, however, several hopeful prospects of
discovering observational signatures that will indeed probe the
strong-gravity domain.
\end{abstract}

\section{Introduction}
     It's fitting to start with a text from Chandrasekhar (1975): ``In
my entire scientific life ....... the most shattering experience has
been the realisation that an exact solution of Einstein's equations of
general relativity, discovered by the New Zealand mathematician Roy
Kerr, provides the absolutely exact representation of untold numbers
of massive black holes that populate the Universe.''

When Chandra wrote this, the evidence was still controversial
(see Israel 1996) : belief
in black holes was at least partly an act of faith (defined by St Paul
as `the substance of things hoped for: the evidence of things not
seen'). But observational progress has been remarkable, especially
within the last couple of years.

I shall first address the question: Do massive collapsed objects exist
-- stellar-mass objects in binaries; and supermassive objects in the
centres of galaxies? The evidence points insistently towards the
presence of dark objects, associated with deep gravitational potential
wells; but it does not in itself tell us about the metric in the
innermost region where Newtonian approximations break down.

 The second part of this paper addresses a separate question: Do these
objects have Schwarzschild/Kerr metrics? Several new observational
probes of the strong-field domain close to the hole have recently
become feasible, offering real prospects of crucially testing our
theories of strong-field gravity.

\section{Stellar-mass black hole candidates}
  It was recognised back in the 1960s that X-ray sources in binary systems,
fuelled by the accretion of gas captured from a companion, would be black hole
candidates if they displayed rapid irregular flickering, and if the inferred
mass were too  high for them to be conventional neutron stars.  The likelihood
of such stellar-mass remnants was of course prefigured by Chandra's classic
early work.

    The discovered systems divide into two categories: those where the
companion star is of high mass, of which Cyg X 1 is the prototype; and
the low-mass X-ray binaries (LMXBs), where the companion is typically
below a solar mass.  The LMXBs are sometimes called `X-ray novae',
because they flare up to high luminosities: they plainly have a
different evolutionary history from systems like Cyg X1. The prototype
of this class is A 0600-00, discovered in 1975. At least 5 further
LMXBs have been discovered more recently, and have had their masses
estimated; other galactic X-ray sources are suspected, on
spectroscopic and other grounds, to be in the same category.  The
strongest current candidates are listed in Table 1, adapted from
Charles (1997). Fuller discussion of these systems, and the
evolutionary scenarios that might lead to them, are given by Tanaka
and Lewin (1995) and Wijers (1996).

    None of these black hole candidates displays the kind of regular
period that is associated, in other systems, with a neutron star's
spin rate. Indeed it is gratifying that, as discussed in John
Freedman's contribution, the putative neutron stars all have masses
clustering around $1.4 M_\odot$, and there are no regularly-pulsing X-ray
sources with dynamically-inferred masses much higher than this. However,
some of the high-mass sources display interesting quasi-periodicities
which (as discussed later) may offer probes of the metric.

    Of course the only black holes that manifest themselves as
conspicuous X-rays sources are the tiny and atypical fraction located
in close binaries where mass transfer is currently going on. There may
be only a few dozen such systems in our Galaxy. However, there is
every reason to suspect that the total number of stellar-mass holes is
at least $10^7$. This is based on the rather conservative estimate
that only one or two percent of supernovae leave black holes rather
than neutron stars. Still larger numbers of holes could indeed exist
(maybe even in the Galactic Halo) as relics of early galactic history.

\section{Supermassive holes}
    Even at the first `Texas' conference, held in 1963 when quasars
had just been discovered, some theorists were suggesting that
gravitational energy, released by a supermassive object, was
responsible for the powerful emitted radiation. There has been a huge
(and increasingly systematic) accumulation of data on quasars, and on
the other classes of active galaxies which are now recognised as being
related: the Seyfert galaxies, already noted as a distinctive category
50 years ago, and the strong radio galaxies (known since the 1950s).
But our understanding has developed in fits and starts. Even if a
convincing explanation has eventually emerged, it sometimes seems as
though this happened only after every other possibility had been
exhausted.

    Recent progress in the study of active galactic nuclei (AGNs)
brings into sharper focus the question of how and when supermassive
black holes formed, and how this process relates to galaxy
formation. Even more important has been the discovery of ordinary
galaxies with equally large redshifts: these were until recently too
faint to be detected, but can now be studied with the combined
resources of the Hubble Space Telescope (HST) and the Keck 10-metre
telescope.  But the most clear-cut and quantitative clues have come
from studies of relatively nearby galaxies: the centres of most of
these display either no activity or a rather low level, but most seem
to harbour dark central masses.  I shall summarise this evidence, and
then outline how it fits in with the broad picture of galaxy formation
and evolution that is now coming into focus.

\subsection{Evidence from the stellar cusp -- M31 (Andromeda) and others}
     Central dark masses -- dead quasars -- have been inferred from
studies of the the spatial distribution and velocities of stars in
several nearby galaxies (see Kormendy and Richstone, 1995; Tremaine,
1997; or van den Marel, 1996 for recent reviews). There is, for
instance, strong evidence for a mass of about $3 \times  10^7 M_\odot$ in
the centre of Andromeda (M31). Even in such a nearby galaxy as this,
the hole's gravitational effects on surrounding stars are restricted
to the central 2-3 arc seconds of the galaxy's image.
Higher-resolution data from  the post-refurbishment HST should
crucially clarify what is going on in these systems.  A list of the
candidates (as of January 1997) is given in Table 2.

\subsection{M87: low-level activity and a supermassive hole}
Although M87, in the Virgo Cluster, was the first in which a
tightly-bound stellar cusp was claimed (Sargent \etal 1978; Young 
\etal{,} 1978), the stellar-dynamics in the core of this giant elliptical
galaxy even now remains rather ambiguous.  According to Merritt and Oh
(1997) the data are consistent with a central mass of $(1-2) 10^9
M_\odot$, but the radial dependence of the projected densities and
velocities could be accounted for by a dense stellar core alone,
provided that the velocities were suitably anisotropic.  Separate
evidence for a dark central mass comes, however, from a disc of gas,
orbiting in a plane perpendicular to the well-known jet (Ford \etal
1994).  One of the complications in studying the central stars in M87
is that the nucleus is not quiescent, but that the inner part of the
jet emits non-thermal light as well as radio waves.

     X-ray data reveal hot gas pervading M87 itself, as well as in the
surrounding cluster.  If there were a huge central hole, then some of
this gas would inevitably be swirling into it, at a rate that can be
estimated. This accretion would give rise to more conspicuous activity
than is observed if the radiative efficiency were as high as 10 per
cent.  Fabian and Canizares (1988) were the first to highlight this
apparent problem with the hypothesis that elliptical galaxies harbour
massive central holes. Actually, the quiescence is less surprising
because, for low accretion rates, the expected luminosity scales as
$\dot M^2$ rather than as $\dot M$: when accretion occurs at a low
rate, and the viscosity is high enough to ensure that the gas swirls
in quickly (so the densities are low), the radiative efficiency also is
low. The gas inflates into a thick dilute torus,where the kinetic
temperature of the ions is close to the virial temperature. Only a
small fraction of the binding energy gets radiated during the time it
takes for each element of gas to swirl inward and be swallowed.
Bremsstrahlung, in particular, is inefficient in this situation; the
most conspicuous emission may be in the radio band, resulting from
synchrotron emission in the strong magnetic field in the inner part of
the accretion flow.  The radio and X-ray emission is actually fully
consistent with accretion at the expected rate, so the observed
non-stellar output from M87 actually corroborates the other evidence
for a supermassive hole. (Fabian and Rees, 1995; Narayan and Yi 1995;
Reynolds \etal 1996; Mahadevan 1997; and references cited therein).

   Weak central radio sources are found in the centres of surprisingly many
otherwise quiescent  ellipticals (Sadler \etal 1995 and references cited
therein). If these galaxies all harbour massive holes, this emission could
similarly be  attributed to accretion in a slow, inefficient, mode (Fabian and
Rees 1995)
 
\subsection{The remarkable case of NGC 4258}
    Much the most compelling case for a central black hole has been
supplied by a quite different technique: amazingly precise mapping of
gas motions via the 1.3 cm maser-emission line of H$_2$O in the peculiar
spiral galaxy NGC 4258 (Watson and Wallin 1994; Miyoshi \etal 1995)
which lies at a distance of about 6.5 Mpc.  The spectral resolution in
the microwave line is high enough to pin down the velocities with
accuracy of 1 km/sec.  The Very Long Baseline Array achieves an
angular resolution better than 0.5 milliarc seconds (100 times sharper
than the HST, as well as far finer spectral resolution of
velocities!).  These observations have revealed, right in the galaxy's
core, a disc with rotational speeds following an exact Keplerian law
around a compact dark mass.  The inner edge of the observed disc is
orbiting at 1080 km/sec. It would be impossible to circumscribe,
within its radius, a stable and long-lived star cluster with the
inferred mass of $3.6 \times 10^7 M_\odot$. The circumstantial
evidence for black holes has been gradually growing for 30 years, but
this remarkable discovery clinches the case completely.  The central
mass must either be a black hole or something even more exotic.

   NGC 4258 poses several puzzles. What determines the sharp inner
edge of the `masing' disc? What is the significance of its inferred
tilt and warping?  How does this thin disc relate to the (thicker)
`molecular tori' that have been postulated in Seyfert galaxies? All
these questions deserve study. It would help, of course, if other
similar discs could be found. Another Seyfert galaxy, NGC 1068, may
show resemblances, but NGC 4258 may prove to be an unusually
fortunate example, because its disc is viewed almost edge-on.
 
\subsection{Our Galactic Centre}
    Most nearby large galaxies seem to harbour massive central holes,
so our own would seem underendowed if it did not have one too. There
has been theoretical advocacy of this view for many years (eg
Lynden-Bell and Rees 1971). Also, an unusual radio source has long
been known to exist right at the dynamical centre of our Galaxy, which
can be interpreted in terms of accretion onto a massive hole (Rees
1982; Melia 1994; Narayan, Yi and Mahadevan 1995). But the direct
evidence has until recently been ambiguous (see Genzel,Townes and
Hollenbach, 1995 and earlier work reviewed therein). This is because
intervening gas and dust in the plane of the Milky Way prevent us from
getting a clear optical view of the central stars, as we can in, for
instance, M31.  A great deal is known about gas motions, from radio
and infrared measurements, but these are hard to interpret because gas
does not move ballistically like stars, being vulnerable to pressure
gradients, stellar winds, and other non-gravitational influences.

    The situation has, however, been transformed by remarkable
observations of stars in the near infrared band, where obscuration by
intervening material is less of an obstacle (Eckart and Genzel
1996). These observations have been made using an instrument (ESO's
`New Technology Telescope' in Chile) with sharp enough resolution to
detect the transverse (`proper') motions of some stars over a
three-year period.  The radial velocities are also known, from
spectroscopy, so one has full three-dimensional information on how the
stars are moving within the central 0.1 pc of our Galaxy. The speeds,
up to 2000 km/sec, scale as $r^{-1/2}$ with distance from the centre,
consistent with a hole of mass $2.5 \times 10^6 M_\odot$.

      In my opinion our Galactic Centre now provides the most
convincing case for a supermassive hole, with the single exception of
NGC 4258.

\subsection{The cumulative evidence} 
A summary of the current evidence is given in Table 2. The data here
(as in table 1) are developing rapidly, and the list may well be longer
by the time this paper appears in print.

    A feature of the data in this table, emphasised by Kormendy and
Richstone (1995) and by Faber \etal (1997), is a crude proportionality
between the hole's mass and that of the central bulge or spheroid in
the stellar distribution (which is of course the dominant part of an
elliptical galaxy, but only a subsidiary component of a disc system
like M31 or our own Galaxy.) This conclusion is only tentative, being
vulnerable to various selection effects, but it suggests that the hole
may form at the same time as the central stellar population. 
In section 5 
I shall
briefly discuss formation scenarios for the holes, in the context of
the remarkable recent progress achieved by optical astronomers in
probing the era when galaxies were still forming.

\section{The fate of stars near a supermassive hole}
\subsection{Tidal disruption}
     Even if a galaxy's core were swept so clean of gas that no
significant emission resulted from steady accretion, there is a
separate process that would inevitably, now and again, liberate a
large supply of gas whenever a supermassive hole was present: tidal
disruption of stars on nearly radial orbits.  A rough estimate, based
on models for the stellar distribution and velocities, suggests that
in M31 a main-sequence star would pass close enough to the putative
hole to be tidally disrupted about once every $10^4$ years.  These
estimates, for M31 and other nearby galaxies, should firm up when
post-refurbishment HST data are available: it is a stellar-dynamical
(rather than gas-dynamical) problem, and therefore relatively `clean'
and tractable.
    
     What happens to a star when it is disrupted?  Earlier
investigations by, for instance, Lacy, Townes and Hollenbach 1982;
Rees 1988; Evans and Kochanek 1989; Canizzo, Lee and Goodman 1990 are
now being supplemented by more detailed numerical modelling (eg
Khokhlov, Novikov and Pethick 1993; Frohlov \etal 1994; Deiner \etal
1997). The tidally disrupted star, as it moves away form the hole,
develops into an elongated banana-shaped structure, the most tightly
bound debris (the first to return to the hole) being at one end (Evans
and Kochanek 1989; Laguna \etal 1993; Kochanek, 1994, Rees 1994).
There would not be a conspicuous `prompt' flare signalling the
disruption event, because the thermal energy liberated is trapped
within the debris.  Much more radiation emerges when the bound debris
(by then more diffuse and transparent) falls back onto the hole a few
months later, after completing an eccentric orbit. The dynamics and
radiative transfer are then even more complex and uncertain than in
the disruption event itself, being affected by relativistic
precession, as well as by the effects of viscosity and shocks (See
Rees 1994, 1996 and earlier work cited therein)

     The radiation from the inward-swirling debris would be
predominantly thermal, with a temperature of order $10^5$ K; however
the energy dissipated by the shocks that occur during the
circularisation would provide an extension into the X-ray band.  High
luminosities would be attained -- the total photon energy radiated (up
to $10^{53}$ ergs) could be several thousand times more than the
photon output of a supernova, though the bolometric correction could
be much larger too. The flares would, moreover, not be standardised --
what is observed would depend on the hole's mass and spin, the type of
star, the impact parameter, and the orbital orientation relative to
the hole's spin axis and the line of sight; perhaps also on absorption
in the galaxy. To compute what happens involves relativistic gas
dynamics and radiative transfer, in an unsteady flow with large
dynamic range, which possesses no special symmetry and therefore
requires full 3-D calculations -- a worthy computational challenge to
those who have many gigaflops at their disposal.

\subsection{Can the `flares' be detected?}
   Supernova-type searches with $10^4$ galaxy-years of exposure should
either detect flares due to this phenomenon, or else place limits on
the mean mass of central black holes in nearby galaxies. This possible
bonus should be an added incentive for such searches.  It is not clear
whether the best strategy involves monitoring nearby galaxies over a
large area of sky or larger numbers of more remote galaxies.  Large
numbers of distant galaxies are, for instance, being routinely
monitored by S. Perlmutter and colleagues in programmes aimed at
discovering supernovae at redshifts of order 0.5.  It would be
surprising if such programmes did not detect such flares -- a negative
result will itself be interesting. However, if a `flare' (with the
expected duration of months) happened in a distant galaxy, one would
not be able to check just how quiescent the galaxy had previously
been. It would be easier to be sure that a detected flare was actually
due to a disrupted star (and not just an upward fluctuation in the
gaseous accretion rate) if it were observed in a closer galaxy that
was known to have previously been inactive.
 
    There has already been possible serendipitous detection of one
transient event in the nucleus of a galaxy (Renzini \etal 1995),
though its peak luminosity was far below what might be expected. X-ray
surveys may also detect the events if (like AGNs) their spectra, though
peaking in the UV, display a high-energy tail (Sembay and West
1993). The predicted flares offer a robust diagnostic of the massive
holes in quiescent galaxies.

\subsection{`Fossil events' in our Galactic Centre?} 
    The rate of tidal disruptions in our Galactic Centre would be no
more than once per $10^5$ years.  But each such event could generate a
luminosity several times $10^{44}$ erg/s for about a year. Were this
in the UV, the photon output, spread over $10^5$ years, could exceed
the current ionization rate: the mean luminosity of the Galactic
Centre might exceed the median value.
    
The resultant fossil ionization would set a lower limit to the
electron density. The radiation emitted from the event might reach us
after a delay if it were reflected off surrounding material. Churazov
\etal (1994) have already used this argument to set a non-trivial
constraint on the history of the Galactic Centre's X-ray output over
the last few thousand years.  Half the debris from a disrupted star
would be ejected on hyperbolic orbits in a fan (which may intersect an
orbiting disc in a line). The structure in the central 2 pc could be a
single spiral feature (Lacy 1994). One speculative possibility (Rees
1987) is that this feature may be a `vapour trail' created by such an
event.

\section{AGN demography and black hole formation}
    The quasar population peaks at redshifts between 2 and 3, but
genuinely seems to be `thinning out' at higher redshifts,
corresponding to still earlier epochs: the comoving density of quasars
falls by at least 3 for each unit increase in $z$ beyond 3 (see
Shaver, 1995 for a review).  The impressive complementary strengths of
HST and the Keck Telescope have revealed galaxies with the same range
of high redshifts as the quasar population itself.  Many of the faint
smudges visible in the Hubble Deep Field (Williams \ etal 1996), the
deepest picture of the sky ever obtained, are galaxies with redshifts
of order 3, being viewed at (or even before) the era when their
spheroids formed.

   Considerations of AGN `demography', by now well known, suggest that
the ultraluminous quasar phase may have a characteristic lifetime set
by the `Eddington timescale' of $4 \times 10^7$ years, being
associated with the formation of a black hole, or the immediate
aftermath of this process.  Straightforward arithmetic based on the
observed numbers of quasars then implies (albeit with substantial
numerical uncertainty because of the poorly known luminosity function,
etc.) that most large galaxies could indeed have gone through a quasar
phase; they would, in consequence, by $z = 2$ ($2-3$ billion years)
have developed central holes of $10^6 -10^9 M_\odot$.

    Physical conditions in the central potential wells, when galaxies
were young and gas-rich, should have been propitious for black hole
formation. Infalling primordial gas would gradually condense into
stars, forming the central spheroid of such systems. But star
formation would be quenched when the gas reached some threshold
central concentration: as the gas evolved (through loss of energy and
angular momentum) to higher densities and more violent internal
dissipation, radiation pressure would inevitably puff it up and
inhibit further fragmentation (Rees, 1993, Haehnelt and Rees
1993). Much of whatever gas remains at this stage would then
agglomerate into a massive hole.

       This argument can be quantified, at least in a crudely
approximate way. A differentially rotating self-gravitating gas mass
can dissipate its energy (via non-axisymmetric instabilities) on a
dynamical timescale. Its internally-generated luminosity can then be
expressed in terms of its virial velocity $v = (GM/r)^{1/2}$ as
$$L = v^5/G = 10^{59} (v/c)^5 \ergps \eqno(1)$$ 
Note the
straightforward analogy to the `maximal power' $c^5/G$ familiar to
relativists and gravitational wave experimenters. This luminosity
reaches the Eddington limit when $v$ is high enough: the gas is then
`puffed up' by radiation pressure, and fragmentation is no longer
possible. The criterion is
$$v > 300 (M/10^6 M_\odot)^{1/5} \kmps\eqno(2)$$ 
This criterion may
be fulfilled for the entire gas mass, or for the inner part of a
self-gravitating disc.  Moreover, while sufficient, (2) is by no means
necessary: fragmentation may be inhibited at a substantially earlier
stage by the effects of higher opacity than electron scattering alone
provides, or by magnetic stresses. If fragmentation is inhibited,
collapse to a supermassive black hole seems almost inevitable.  To
evade such an outcome, either:\par 
(i) Stars must form (before (2) is
satisfied) with nearly 100 per cent efficiency; moreover, they must
all have low mass (so that no material is expelled again) {\it or}

(ii) 
Gas must remain in a self-gravitating disc for hundreds of orbital
periods, without the onset of any instability that redistributes
angular momentum and allows the inner fraction to collapse enough to
cross the threshold when (2) applies.

    Neither of these `escape routes' seems at all likely -- the first
would require the stars to have an initial mass function quite
different from what is actually observed in the spheroids of galaxies;
the second is contrary to well-established arguments that
self-gravitating discs are dynamically unstable.  The mass of the hole
would depend on that of its host galaxy, though not necessarily via an
exact proportionality: the angular momentum and the depth of the
galaxy's potential well are relevant factors too.

   This process involves complex gas dynamics and feedback from stars;
we are still a long way from being able to make realistic
calculations. At the moment, the most compelling argument that a
massive black hole is an expected byproduct comes from the
implausibility of the alternatives.  The mass of the hole would depend
on that of its host galaxy, though not necessarily via an exact
proportionality: the angular momentum of the protogalaxy and the depth
of its central potential well are relevant factors too. A more
quantitative estimate depends on calculating in full detail when,
during the progressive concentration towards the centre, star
formation ceases (because of radiation pressure, magnetic fields, or
whatever) and the remaining gas evolves instead into a supermassive
object.
 
    Once a large mass of gas became too condensed to fragment into
stars, it would continue to contract and deflate. Some mass would
inevitably be shed, carrying away angular momentum, but the remainder
could continue contracting until it underwent complete gravitational
collapse. This could be a substantial fraction -- for example, if 10
per cent of the mass had to be shed in order to allow contraction by a
factor of 2, about 20 per cent could form a black hole.

  Firmer and more quantitative conclusions will have to await
elaborate numerical simulations. But on one issue I would already bet
strongly. This is that a massive black hole forms directly from gas
(some, albeit, already processed through stars), perhaps after a
transient phase as a supermassive object, rather than from coalescence
of stars or mergers of stellar-mass holes.

       The energy radiated during further growth of the hole manifests
itself as a quasar. The peak in the quasar population (i.e. redshifts
in the range 2 - 3) signifies the era when large galactic spheroids
were forming in greatest profusion. It is worth noting, incidentally,
that whereas activity in low-z galaxies may be correlated with some
unusual disturbance due to a tidal encounter or merger, this may not
be the right way to envisage the more common high-z quasars. Any
newly-formed galaxy is inevitably `disturbed', in the sense that it has
not yet had time to settle down and relax: no external influence is
needed to perturb axisymmetry or to trigger a large inflow of gas.

\section{Do the candidate holes obey the Kerr metric?}
\subsection{Probing the region near the hole}
     As already discussed in section 3, NGC 4258 offers the clearest
evidence so far for a central dark mass. But the observed molecular
disc lies a long way out: at around $10^5$ gravitational radii.  We can
exclude all conventional alternatives (dense star clusters, etc);
however, the measurements tell us nothing about the central region
where gravity is strong, certainly not whether the putative hole
actually has properties consistent with the Kerr metric. The stars in
the central parts of M31 and our own galaxy likewise lie so far out
that their orbits are essentially Newtonian.

        The phenomena of AGNs are due to material closer to the
central mass, but nobody could yet claim that any observed features of
AGNs offers a clear diagnostic of a Kerr metric. All we can really
infer is that `gravitational pits' exist, which must be deep enough to
allow several percent of the rest mass of infalling material to be
converted into kinetic energy, and then radiated away from a region
compact enough to vary on timescales as short as an hour. General
relativity has been resoundingly vindicated in the weak field limit
(by high-precision observations in the Solar System, and of the binary
pulsar) but we still lack quantitative probes of the strongly
relativistic region.
   
    The tidal disruption events described in section 4 depend
crucially on distinctive precession effects around a Kerr metric, but
the gas dynamics are so complex and messy that even when a flare is
detected it will not serve as a useful diagnostic of the metric in the
strong-field domain. On the other hand, the stars whose motions reveal
a central dark mass in our Galactic Centre, in M31, and in other
normal galaxies are in orbits $\simgt 10^5$ times larger than the
putative holes themselves.

    Relativists would seize eagerly on any relatively `clean' probe of
the relativistic domain.  In most accretion flows, the emission is
concentrated towards the centre, where the potential well is deepest
and the motions fastest.  Such basic features of the phenomenon as the
overall efficiency, the minimum variability timescale, and the
possible extraction of energy from the hole itself all depend on
inherently relativistic features of the metric -- on whether the hole
is spinning or not, how it is aligned, etc. There are now several
encouraging new possibilities.
 
\subsection{X-ray  spectroscopy of accretion flows}
       Optical spectroscopy tells us a great deal about the gas in
AGNs.  However, the optical inferences pertain to gas that is quite
remote from the hole itself. This is because the innermost regions
would be so hot that their thermal emission emerged as more energetic
quanta: the optical observations sample radiation that is emitted (or
at least reprocessed) further out.  The X-rays, on the other hand,
come predominantly from the relativistic region.  Until recently,
however, the energy resolution and sensitivity of X-ray detectors was
inadequate to permit the study of line shapes. But this is now
changing.  The ASCA X-ray satellite was the first to offer sufficient
spectral resolution to reveal line profiles, and therefore opened up
the possibility of seeking the substantial gravitational redshifts, as
well as large doppler shifts, that would be expected. (Fabian \etal
1989, and earlier references cited therein).  There is already one
convincing case (Tanaka \etal 1995) of a broad asymmetric emission
line indicative of a relativistic disc viewed at $\sim 60^\circ$ to its
plane, and others are now being found.  The value of (a/m) can in
principle be constrained too, because the emission is concentrated
closer in, and so displays larger shifts, if the hole is rapidly
rotating (Iwasawa \etal 1996).

  The appearance of a disc around a hole, taking doppler and
gravitational shifts into account, along with light bending, was
calculated by Bardeen and Cunningham (1973) and by several other
authors. The associated swing in the polarization vector of photon
trajectories near a hole was also long ago suggested (Connors, Piran
and Stark 1980) as another diagnostic; but this is still not feasible
because X-ray polarimeters are far from capable of detecting the few
per cent polarization expected.
 
\subsection{Stars in relativistic orbits?}
   These X-ray observations are of course of Seyfert galaxies, whose
centres, though not emitting as powerfully as quasars, are by no means
inactive.  But we still need a `cleaner' and more quantitative probe
of the strong-field regime.

    A small star orbiting close to a supermassive hole would behave
like a test particle, and its precession would probe the metric in the
`strong field' domain.  These interesting relativistic effects, have
been computed in detail by Karas and Vokrouhlicky (1993, 1994) and
Rauch (1997).  Would we expect to find a star in such an orbit?

   An ordinary star certainly cannot get there by the kind of `tidal
capture' process that can create close binary star systems. This is
because the binding energy of the final orbit (a circular orbit with
radius $2r_T$, which has the same angular momentum as an initially
near-parabolic orbit with pericentre at $r_T$) is far higher when the
companion is a supermassive hole than when it is also of stellar mass
-- it scales roughly as $M^{2/3}$. This orbital energy would have to
be dissipated within the star, and that cannot happen without
destroying it: a star whose orbit brings it within (say) $3r_T$ of a
massive black hole may not be destroyed on first passage (as described
in section 4); however, is then on a bound elliptical orbit, it will
surely be disrupted before the orbit has circularised. (It would then
give a `flare' similar to that discussed in section 4, but with a
somewhat longer timescale.)

    Syer, Clarke and Rees (1991) pointed out, however, that an orbit
can be `ground down' by successive impacts on a disc (or any other
resisting medium) without being destroyed: the orbital energy then
goes almost entirely into the material knocked out of the disc, rather
than into the star itself. Other constraints on the survival of stars
in the hostile environment around massive black holes -- tidal
dissipation when the orbit is eccentric, irradiation by ambient
radiation, etc -- are explored by Podsiadlowski and Rees (1994), and
King and Done (1993).
 
   These stars would not be directly observable, except maybe in our
own Galactic Centre.  But they might have indirect effects: such a
rapidly-orbiting star in an active galactic nucleus could signal its
presence by quasiperiodically modulating the AGN emission.
 
    There was a flurry of interest some years ago when X-ray
astronomers detected an apparent 3.4 hour periodicity in the Seyfert
galaxy NGC 6814.  But it turned out that there was a foreground binary
star, with just that period, in the telescope's field of view. But
theorists shouldn't be downcast. It is more elevated to make
predictions than to explain phenomena a posteriori, and that's all we
can now do. There is a real chance that someday observers will find
evidence that an AGN is being modulated by an orbiting star, which
could act as a test particle whose orbital precession would probe the
metric in the domain where the distinctive features of the Kerr
geometry should show up clearly.

\subsection{Gravitational-wave capture of compact stars} 
   Objects circling close to supermassive black holes could be neutron
stars or white dwarfs, rather than ordinary stars. Such compact stars
would be impervious to tidal dissipation, and would have such a small
geometrical cross section that the `grinding down' process would be
ineffective too. On the other hand, because they are small they can
get into very tight orbits by straightforward stellar-dynamical
processes. For ordinary stars, the `point mass' approximation breaks
down for encounter speeds above 1000 km/s -- physical collisions are
then more probable than large-angle deflections: but there is no
reason why a `cusp' of tightly bound {\it compact} stars should not
extend much closer to the hole.  Neutron stars or white dwarfs could
exchange orbital energy by close encounters with each other until some
got close enough that they either fell directly into the hole, or
until gravitational radiation became the dominant energy loss.
Gravitational radiation losses  tend to circularise an elliptical
orbit with small pericentre. Most stars in such orbits would be
swallowed by the hole before circularisation, because the angular
momentum of a highly eccentric orbit `diffuses' faster than the energy
does due to encounters with other stars, but some would get into close
circular orbits (Hils and Bender 1995; Sigurdsson and Rees 1997).

    A compact star is less likely than an ordinary star in similar
orbit to `modulate' the observed radiation in a detectable way.  But
the gravitational radiation (almost periodic because the dissipation
timescale involves a factor $(M/m*)$) might eventually be detectable
(see below).

\subsection{The Blandford-Znajek process}
   Blandford and Znajek (1977) showed that a magnetic field threading
a hole (maintained by external currents in, for instance, a torus)
could extract spin energy, converting it into directed Poynting flux
and electron-positron lairs.  This is, in effect, an
astrophysically-realistic example of the Penrose (1969) process
whereby the spin of a Kerr hole can be tapped. It would indeed be
exciting if we could point to objects where this was happening. The
centres of galaxies display a bewildering variety of phenomena, on
scales spanning many powers of 10. The giant radio lobes sometimes
spread across millions of lightyears -- $10^{10}$ times larger than the
hole itself. If the Blandford-Znajek process is really going on (Rees
\etal 1982) these huge structures may be the most direct
manifestation of an inherently relativistic effect around a Kerr hole.

    Jets in some AGNs definitely have Lorentz factors $\gamma_j$
exceeding 10. Moreover, some are probably Poynting-dominated, and
contain pair (rather than electron-ion) plasma. But there is still no
compelling reason to believe that these jets are energised by the hole
itself, rather than by winds and magnetic flux `spun off' the
surrounding torus. The case for the Blandford-Znajek mechanism would
be strengthened if baryon-free jets were found with still higher
$\gamma_j$ , or if the spin of the holes could be independently
measured, and the properties of jets turned out to depend on $(a/m)$.

\subsection{Scaling laws and `microquasars'}
Two of the galactic X-ray sources that are believed to involve black
holes (See Table 1) generate double radio structures that resemble
miniature versions of the classical extragalactic strong radio
sources. The jets have been found to display apparent superluminal
motions across the sky, indicating that, like the extragalactic radio
sources, they contain plasma that is moving relativistically (Mirabel
and Rodriguez 1994).

  There is no reason to be surprised by this analogy between phenomena
on very different scales. Indeed, the physics of flows around black
holes is always essentially the same, apart from very simple scaling
laws. If we define $l = L/L_{Ed}$ and $\dot m = \dot M/\dot M_{crit}$,
where $\dot M_{crit} = L_{Ed}/c^2$, then for a given value of $\dot
m$, the flow pattern may be essentially independent of $M$. Linear
scales and timescales, at a given value of $r/r_g$, where $r_g =
GM/c^2$, are proportional to $M$, and densities in the flow for a
given $\dot m$ then scale as $M^{-1}$.  The physics that amplifies and
tangles any magnetic field may be scale-independent; the field
strength $B$ then scales as $M^{-1/2}$. So the bremsstrahlung or
synchrotron cooling timescales (proportional to $\rho ^{-1}$ and
$B^{-1/2}$ respectively) go as $M$, implying that $t_{cool}/t_{dyn}$
depends primarily on $\dot m$. So also do the  ratios involving,
for instance, coupling of electron and ions in thermal
plasma. Therefore, the efficiencies and the value of $l$ are
insensitive to $M$, and depend primarily on $\dot m$.  Moreover, the form
of the spectrum depends on $M$ only rather insensitively (and in a
manner that is easily calculated).

  The kinds of accretion flow inferred in, for instance, M87, giving
rise to a compact radio and X-ray source, along with a relativistic
jet, could operate just as well if the hole mass was lower by a
hundred million, as in the galactic LMXB sources.  So we can actually
study the processes involved in AGNs in microquasars close at hand
within our own galaxy. And we may even be able to see the entire
evolution of a strong extragalactic radio source, speeded up by a
similar factor.

\subsection{Discoseismology}
   Discs or tori that are maintained by steady flow into a black hole
can support vibrational modes (Kato and Fukui 1980; Nowak and Wagoner
1992, 1993).  The frequencies of these modes can, as in stars, serve as
a probe for the structure of the inner disc or torus. The amplitude
depends on the importance of pressure, and hence on disc thickness;
how they are excited, and the amplitude they may reach, depends, as in
the Sun, on interaction with convective cells and other macroscopic
motions superimposed on the mean flow. But the {\it frequencies} of
the modes can be calculated more reliably. In particular, the lowest
g-mode frequency is close to the maximum value of the radial epicyclic
frequency k.  This epicyclic frequency is, in the Newtonian domain,
equal to the orbital frequency. It drops to zero at the innermost
stable orbit. It has a maximum at about $9GM/c^2$ for a Schwarzschild
hole; for a Kerr hole, k peaks at a smaller radius (and a higher
frequency for a given $M$). The frequency is 3.5 times higher for
$(a/m)=1$ than for the Schwarzschild case.

   Nowak and Wagoner pointed out that these modes may cause an
observable modulation in the X-ray emission from galactic black hole
candidates.  Just such effects have been seen in GRS 1915+105 (Morgan
\etal 1996). The amplitude is a few per cent (and somewhat larger at
harder X-ray energies, suggesting that the oscillations involve
primarily the hotter inner part of the disc). The fluctuation spectrum
shows a peak in Fourier space at around 67 Hz. This frequency does not
change even when the X-ray luminosity doubles, suggesting that it
relates to a particular radius in the disc. If this is indeed the
lowest g-mode, and if the simple disc models are relevant, then the
implied mass is $10.2 M_\odot$ for Schwarzchild, and $35 M_\odot$ for
a `maximal Kerr' hole (Nowak \etal 1997).  The mass of this system is
not well known. However, this technique offers the exciting prospect
of inferring $(a/m)$ for holes whose masses are independently known.

    GRS 1915+105 is one of the objects with superluminal radio
jets. The simple scaling arguments of section 6.6 imply that the AGNs
which it resembles might equally well display oscillations with the
same cause. However, the periods would be measured in days, rather
than fractions of a second.

\section{Gravitational radiation as a probe}
\subsection{Gravitational waves from newly-forming massive holes?}
     The gravitational radiation from black holes, as Kip Thorne's
paper emphasises, offers potentially impressive tests of general
relativity, involving no physics other than the dynamics of spacetime
itself.
  
At first sight, the original formation of the holes might seem the
most obvious sources of strong wave pulses.  However the wave emission
would only be efficient if the holes formed on a timescale as short as
($r_g/c$)  -- something that might happen if
they built up via coalescence of smaller holes (cf Quinlan and Shapiro
1990).

If, on the other hand, supermassive black holes formed as suggested in
section 5 -- directly from gas (some, albeit, already processed
through stars), perhaps after a transient phase as a supermassive
object -- then the process would be too gradual to yield efficient
gravitational radiation. The least pessimistic scenario from the
perspective of gravitational-wave astronomers, in the context of these
latter ideas, would be one in which a supermassive star accumulates,
and then collapses into a hole, on a dynamical timescale, via
post-Newtonian instability. But even this yields much weaker
gravitational radiation than black hole coalescence. That is because
post-Newtonian instability is triggered at a radius $r_i \gg r_g$.
Supermassive stars are fragile because of the dominance of radiation
pressure: this renders the adiabatic index $\Gamma$ only slightly
above 4/3 (by an amount of order $(M/M_\odot )^{-1/2}$).  Since
$\Gamma = 4/3$ yields neutral stability in Newtonian theory, even the
small post-Newtonian corrections then destabilise such
`superstars'. The characteristic collapse timescale when instability
ensues is longer than $r_g/c$ by the 3/2 power of that factor, and the
total gravitational wave energy emitted is lower by the
cube. Efficiency might be enhanced if the specific angular momentum
when the instability occurred were just above the limit that could be
accepted by a newly-formed Kerr hole.  Material falling inward would
then accumulate in a disc or pancake structure with dimensions only a
few times $r_g$; if ordinary viscosity were ineffective in expelling
the excess angular momentum, the disc might then become sufficiently
assymetric that gravitational waves could do the job.

      If the material were initially of uniform density (a `top hat'
distribution) and then fell in freely, it would all reach the centre
simultaneously. However, this would not happen if the pre-collapse
density profile were characteristic of a supermassive object.
Different shells of material would reach the centre at times spread by
roughly the initial free-fall time, larger than ($r_g/c)$ by a factor
($r_i/r)^{3/2}$ . (In the spherical case, $r_i/r_g$ would be
$(M/M_\odot )^{1/2}$).  If other mechanisms for angular momentum transfer
could be suppressed, the resultant ring of material would swirl
inward, owing to loss of angular momentum via gravitational radiation,
in $(M_{ring}/M)^{-1}$ orbital periods.  A quasi-steady state could
therefore be maintained for the overall-free fall timescale $\sim
(r_i/r)^{3/2} (r_g/c)$, during which material drains inward so that
the amount stored in the ring maintains itself at $(r_i/r)^{-3/4} M$.
The gravitational radiation would be `efficient' in the sense that it
carried away a significant fraction of the rest-mass energy, but this
would happen over a longer period than $r_g/c$, so the amplitude would
be lower by $(r_i/r)^{-3/4}$.  (I should emphasise that this example is
merely illustrative, and is obviously not very realistic.)

The important point is that the formation of a hole `in one go' from a
supermassive star is an unpromising source of gravitational waves. If
the hole grows more gradually, then the prospects are obviously still
worse. On the other hand, if the host galaxy had not yet acquired a
well-defined single centre, several separate holes could form, and
yield strong events when they subsequently coalesce.

    The gravitational waves associated with supermassive holes would
be concentrated in a frequency range around a millihertz -- too low to
be accessible to ground-based detectors, which lose sensitivity below
100 Hz, owing to seismic and other background noise. Space-based
detectors are needed.  One such, proposed by the European Space
Agency, is the Laser Interferometric Spacecraft (LISA) -- six
spacecraft in solar orbit, configured as two triangles, with a
baseline of 5 million km whose length is monitored by laser
interferometry.
    
\subsection{Coalescing supermassive holes.}
     The guaranteed sources of really intense gravitational waves in
LISA's frequency range would be coalescing supermassive black
holes. Many galaxies have experienced a merger since the epoch $z > 2$
when, according to `quasar demography' arguments (section 3) they
acquired central holes.  The holes in the two merging galaxies would
spiral together, emitting, in their final coalescence, up to 10 per
cent of their rest mass as a burst of gravitational radiation in a
timescale of only a few times $r_g/c$. These pulses would be so strong
that LISA could detect them with high signal-to-noise even from large
redshifts.  Whether such events happen often enough to be interesting
can to some extent be inferred from observations (we see many galaxies
in the process of coalescing), and from simulations of the
hierarchical clustering process whereby galaxies and other cosmic
structures form.  Haehnelt (1994) calculated the merger rate of the
large galaxies believed to harbour supermassive holes: it is only
about one event per century, even out to redshifts $z = 4$. Mergers of
small galaxies are more common -- indeed big galaxies are probably the
outcome of many successive mergers. We have no direct evidence on
whether these small galaxies harbour black holes (nor, if they do, of
what the hole masses typically are). However it is certainly possible
that enough holes of (say ) $10^5 M_\odot$ lurk in small early-forming
galaxies to yield, via subsequent mergers, more than one event per
year detectable by LISA.

\subsection{Effects of recoil}
   There would be a recoil due to the non-zero net {\it linear}
momentum carried away by gravitational waves in the coalescence. If
the holes have unequal masses, a preferred longitude in the orbital
plane is determined by the orbital phase at which the final plunge
occurs.  For spinning holes there may be a rocket effect perpendicular
to the orbital plane, since the spins break the mirror symmetry with
respect to this  plane. (Redmount and Rees, 1989 and references
cited therein.)

   The recoil is a strong-field gravitational effect which depends
essentially on the lack of symmetry in the system.  It can therefore
only be properly calculated when fully 3-dimensional general
relativistic calculations are feasible.  The velocities arising from
these processes would be astrophysically interesting if they were
enough to dislodge the resultant hole from the centre of the merged
galaxy, or even eject it into intergalactic space.

      LISA is potentially so sensitive that it could detect the
nearly-periodic waves waves from stellar-mass objects orbiting a $10^5
- 10^6 M_\odot$ hole, even at a range of a hundred Mpc, despite the
$m/M$ factor whereby the amplitude is reduced compared with the
coalescence of two objects of comparable mass $M$. The stars in the
observed `cusps' around massive central holes in nearby galaxies are
of course (unless almost exactly radial) on orbits that are far too
large to display relativistic effects.  Occasional captures into
relativistic orbits can come about by dissipative processes -- for
instance, interaction with a massive disc (eg Canizzo, Lee and Goodman
1990; Syer, Clarke and Rees 1991). But unless the hole mass were above
$10^8 M_\odot$ (in which case the waves would be at too low a frequency for
LISA to detect), solar-type stars would be tidally disrupted before
getting into relativistic orbits. Interest therefore focuses on
compact stars, for which dissipation due to tidal effects or drag is
less effective. As described in section 6.3, compact stars may get
captured as a result of gravitational radiation, which can gradually
`grind down' an eccentric orbit with close pericenter passage into a
nearly-circular relativistic orbit (Hils and Bender 1995; Sigurdsson
and Rees 1996). The long quasi-periodic wave trains from such objects,
modulated by orbital precession (cf Karas and Vokrouhlicky 1993; Rauch
1997), in principle carry detailed information about the metric.

     The attraction of LISA as an `observatory' is that even
conservative assumptions lead to the prediction that a variety of
phenomena will be detected.  If there were many massive holes not
associated with galactic centres (not to mention other speculative
options such as cosmic strings), the event rate could be much
enhanced. Even without factoring in an `optimism factor' we can be
confident that LISA will harvest a rich stream of data.
  
     LISA is at the moment just a proposal -- even if funded, it is
unlikely to fly before 2017.  (It will cost perhaps 3 times as much as
LIGO Phase 1, but may detect infinitely more events). Is there any way
of learning, before that date, something about gravitational
radiation?  The dynamics (and gravitational radiation) when two holes
merge has so far been computed only for cases of special symmetry. The
more general problem -- coalescence of two Kerr holes with general
orientations of their spin axes relative to the orbital angular
momentum -- is one of the US `grand challenge' computational
projects. When this challenge has been met (and it will almost
certainly not take all the time until 2017) we shall find out not only
the characteristic wave form of the radiation, but the recoil that
arises because there is a net emission of linear momentum.

    This recoil could displace the hole from the centre of the merged
galaxy (Valtonen 1996 and references therein) -- it might therefore be
relevant to the low-z quasars that seem to be asymmetrically located
in their hosts (and which may have been activated by a recent
merger). Even galaxies that do not harbour a central hole may,
therefore, once have done so in the past. The core of a galaxy that
has experienced such an ejection event may retain some trace of it
(perhaps, for instance, an unusual profile), because of the energy
transferred to stars via dynamical friction during the merger process
(cf Ebisuzaki, Makino and Okumura 1991; Faber \etal 1997).

   The recoil might even be so violent that the merged hole breaks
loose from its galaxy and goes hurtling through intergalactic space.
This disconcerting thought should impress us with the reality and
`concreteness' of the entities whose theoretical properties Chandra did
so much to illuminate.

\section{References}

\ref
Blandford, R.D. and Znajek, R.L. 1977, MNRAS    
{\bf 179}, 433.
\ref
Canizzo, J.K., Lee, H.M. and Goodman, J. 1990,    ApJ    {\bf 351}, 
38.
\ref
Chandrasekhar, S. 1975,    lecture reprinted in `Truth and Beauty'  
(Chicago U.P.
1987) p54.
\ref
Charles, P.A. 1997, Proc 18th Texas Conference 
(World Scientific, Singapore)    (in
press).
\ref
Churazov, E \etal 1994,   ApJ Supp   {\bf  92}, 381.
\ref
Connors, P.A., Piran, T.  and Stark, R.F. 1980, ApJ {\bf 235}, 224. 
\ref
Cunningham, C.T. and Bardeen, J.M. 1993, ApJ {\bf 183}, 237.
\ref
Deiner, P. \etal 1997
\ref
Ebisuzaki, T., Makino, J., and Okumura, S.K. 1991, Nature {\bf 354}, 212.
\ref
Eckart, A.  and Genzel, R., 1996 Nature {\bf 383}, 415.
\ref
Evans, C.R. and Kochanek C.S. 1989, ApJ (Lett) 346, L13.
\ref
Faber, S.M. \etal 1997 ApJ (in press).
\ref
Fabian, A.C. and Canizares, C.R.,  1988, Nature {\bf 333}, 829. 
 \ref
Fabian, A.C. Rees, M.J., Stella, L and White, N.E. 1989, MNRAS {\bf 238}, 729.
\ref
Fabian, A.C. and Rees, M.J. 1995, MNRAS {\bf 277}, L55. 
\ref
Ford, H. C  \etal, 1994, ApJ {\bf 435}, L27.
\ref
Frolov, V.P. \etal 1994, ApJ {\bf 432}, 680.
\ref
Genzel, R., Townes, C.H. and Hollenbach, D.J. 1994,  Rep. Prog. Phys 
{\bf 57}, 417.
 \ref
Haehnelt M 1994, MNRAS {\bf 269},199.
\ref
Haehnelt M, and Rees, M.J. 1993, MNRAS {\bf 263}, 168.
\ref
Israel, W. 1996, Foundations of Physics {\bf 26}, 595.
\ref
Iwasawa, K. \etal 1996, MNRAS {\bf 282}, 1038.
\ref
Hils, D and Bender, P.L.  1995, ApJ (Lett) {\bf 445}, L7. 
\ref
Karas, V., and Vokrouhlicky, D., 1993, MNRAS {\bf 265}, 365.
\ref
Karas, V., and Vokrouhlicky, D., 1994, ApJ {\bf 422}, 208.
\ref
Kato, S   and Fukui, J.  1980, PASJ {\bf 32}, 377. 
\ref
Khokhlov, A., Novikov, I.D. and Pethick, C.J. 1993,  ApJ {\bf 418}, 163.
\ref
King, A.R and Done, C. 1993, MNRAS {\bf 264}, 388
\ref
Kochanek, C.S., 1994, ApJ {\bf 422}, 508
\ref
Kormendy, J. and Richstone, D. 1995, Ann Rev. Astr. Astrophys {\bf 33}, 581.
\ref
Lacy, J.H. 1994 in `The nuclei of normal galaxies' ed R. Genzel and A.I. Harris
(Kluwer) p165.
\ref
Lacy, J.H., Townes, C.H. and Hollenbach, D.J. 1982, ApJ {\bf 262}, 120.
\ref
Laguna, P.,  Miller, W.A., Zurek, W.H., and Davies, M.B. 1993, ApJ
(Lett) {\bf 410}, L83.
\ref
Lynden-Bell, D. and Rees, M.J. 1971, MNRAS {\bf 152}, 461. 
\ref
Mirabel, I.F. and Rodriguez, L.F., 1994,  Nature {\bf 371}, 48.
\ref
Miyoshi, K \etal 1995, Nature {\bf 373}, 127. 
\ref
Mahadevan, R. 1997, ApJ (in press).
\ref
Merritt, D and Oh, S.P. 1997, Astron J (in press).
\ref
Melia, F, 1994, ApJ {\bf 426}, 577.
\ref
Morgan, E., Remillard, R and Greiner, J. 1996, IAU Circular No. 6392
\ref
Narayan, R and Yi, I. 1995, ApJ {\bf 444}, 231.
\ref
Narayan, R., Yi, I, and Mahadevan, R. 1995, Nature {\bf 374}, 623.
\ref
Nowak, M.A.  and Wagoner, R.V. 1992, ApJ {\bf 393}, 697.
\ref
Nowak, M.A.  and Wagoner, R.V. 1993, ApJ {\bf 418}, 187.
\ref
Nowak, M.A., Wagoner, R.V., Begelman, M.C. and Lehr, D.E. 1997 ApJ (in
press). 
\ref
Penrose, R. 1969, Rev. Nuovo. Cim {\bf 1}, 252.
\ref
Podsiadlowski, P. and Rees, M.J. 1994, in `Evolution of X-ray binaries' ed 
Holt and C.Day (AIP) p403.
\ref
Quinlan, G.D. and Shapire, S.L. 1990,  ApJ {\bf 356}, 483.
\ref
Rauch, K. P. 1997, ApJ (in press).
\ref
Redmount, I. and Rees, M.J. 1989, Comm. Astrophys. Sp. Phys {\bf 14}, 185.
\ref
Rees, M.J. 1982 in `The Galactic Center' ed  G. Riegler and R.D. Blandford
(A.I.P) p166.
\ref
Rees, M.J. 1987 in `Galactic Center' eds  Backer, D. and Genzel, R. (AIP
Conference Proceedings).
\ref
Rees, M.J. 1988, Nature {\bf 333}, 523.
\ref
Rees, M.J. 1993, Proc. Nat. Acad. Sci {\bf  90}, 4840.
\ref
Rees, M.J. 1994 in `Nuclei of Normal Galaxies' ed R. Genzel and A.I. Harris
(Kluwer) p453
\ref
Rees, M.J. 1996 in `Gravitational Dynamics' ed O. Lahav \etal (C.U.P.) p 103.
\ref
Rees, M.J., Begelman, M.C., Blandford, R.D. and Phinney, E.S. 1982, Nature 
{\bf 295}, 17.
\ref
Renzini, A \etal 1995, Nature  {\bf 378}, 39.
\ref
Reynolds, C \etal 1996, MNRAS {\bf 283}, L111.
\ref
Sadler, E.M. \etal 1995, MNRAS {\bf 276}, 1373.
\ref
Sargent, W.L.W. \etal 1978, ApJ  {\bf 221}, 731
\ref
Sembay , S and West, R.G. 1993, MNRAS {\bf 262}, 141.
\ref
Shaver, P. 1995, Ann. N.Y. Acad. Sci {\bf 759}, 87.
\ref
Sigurdsson, S and Rees M.J. 1997, MNRAS {\bf 284}, 318.
\ref
Syer, D., Clarke, C.J. and Rees, M.J. 1991, MNRAS {\bf 250}, 505.
\ref
Tanaka, Y and Lewin, W.H.G. 1995 in X-ray Binaries, ed W.H.G. Lewin \etal (CUP)
p126.
\ref
Tanaka, Y \etal 1995, Nature {\bf 375}, 659.
\ref
Tremaine, S 1997 in `Some Unsolved Problems in Astrophysics' eds J.   
   Bahcall and J.P. Ostriker (Princeton U.P.) (in press).
\ref
Valtonen, M. 1996, Comments Astrophys. {\bf 18}, 191.
\ref
van den Marel, R 1996 in `New Light on Galactic Evolution' eds R. Bander
    and R. Davies (Kluwer) (in press).
\ref
Watson, W.D. and Wallin, B.K. 1994, ApJ (Lett) {\bf 432}, L35.
\ref
Wijers, R.A.M.J.  1996 in `Evolutionary Processes in Binary Stars' eds R.A.M.J.
Wijers \etal (Kluwer), p327
\ref
Williams, R \etal 1996, Astron. J. {\bf 112}, 1335.
\ref 
Young, P. J. \etal 1978, ApJ {\bf 221}, 721.
\vfill\eject
\bigskip
\centerline{\bf Table I}
\smallskip
\centerline{\bf Stellar-mass black hole candidates and their binary companions}
\bigskip

\halign{\vtop{\parindent=0pt\parskip=0pt\hsize=1.5truein\strut #
\hfill\strut}&
\vtop{\parskip=0pt\parindent=0pt\hsize=1.5truein\strut\hfill #\hfill\strut}&
\vtop{\parskip=0pt\parindent=0pt\hsize=1.5truein\strut\hfill #\hfill\strut}\cr
&$M_h/M_\odot$&$M_*/M_\odot$\cr
\noalign{\medskip}
{\it High mass companions}\cr
\noalign{\smallskip}
Cyg X1&11-21&24-42\cr
LMC X3&5.6-7.8&20\cr
\noalign{\bigskip\bigskip\noindent {\it Low mass companions 
(X-ray transients)}\medskip}
V404 Cyg&10-15&$\sim 0.6$\cr
A 0620-00&5-17&0.2-0.7\cr
Nova Muscae&4.2-6.5&0.5-0.8\cr
GS 2000+25&6-14&$\sim 0.7$\cr
GROJ1655-40&4.5-6.5&$\sim 1.2$\cr
N. Oph 77&5-9&$\sim 0.4$\cr
J0422432&6-14&$\sim 0.3$\cr}
\bigskip
\bigskip
\bigskip
\centerline{\bf Table II}
\smallskip
\centerline{\bf Supermassive holes}
\bigskip 

\halign{\vtop{\parindent=0pt\parskip=0pt\hsize=1.5truein\strut #
\hfill\strut}&
\vtop{\parskip=0pt\parindent=0pt\hsize=1.5truein\strut\hfill #\hfill\strut}&
\vtop{\parskip=0pt\parindent=0pt\hsize=2.00truein\strut\hfill #\hfill\strut}\cr
&$M_h/M_\odot$&Method\cr
\noalign{\medskip}
M87&$2.10^9$&stars+opt.disc\cr
NGC 3115&$10^9$&stars\cr
NGC 4486 B&$5.10^8$&stars\cr
NGC 4594 (Sombrero)&$5.10^8$&stars\cr
NGC 3377&$8.10^7$&stars\cr
NGC 3379&$5.10^7$&stars\cr
NGC 4258&$4.10^7$&masing H$_2$O disc\cr
M31 (Andromeda)&$3.10^7$&stars\cr
M32&$3.10^6$&stars\cr
Galactic Centre&$2.5. 10^6$&stars+(3-D motions)\cr}

\vfill
\end{document}